\documentclass[osajnl,preprint,showpacs,floatfix]{revtex4}
\usepackage{graphicx}
\usepackage{dcolumn}
\usepackage{amsmath}
\usepackage{here}
\begin{document}

\title{Incoherent vector vortex-mode solitons in self-focusing nonlinear media}

\author{Kristian Motzek and Friedemann Kaiser}

\affiliation{Institute of Applied Physics, Darmstadt University of
Technology, D-64289 Darmstadt, Germany}

\author{Jos\'e R. Salgueiro and Yuri Kivshar}

\affiliation{Nonlinear Physics Center, Research School of Physical
Sciences and Engineering, Australian National University, Canberra
ACT 0200, Australia}

\author{Cornelia Denz}

\affiliation{Nonlinear Photonics Group, Institute of Applied
Physics, Westf\"alische Wilhelms-Universit\"at M\"unster, D-48149
M\"unster, Germany}

\begin{abstract}
We suggest a novel type of composite spatial optical soliton
created by a coherent vortex beam guiding a partially incoherent
light beam in a self-focusing nonlinear medium. We show that the
incoherence of the guided mode may enhance, rather than suppress,
the vortex azimuthal instability, and also demonstrate strong
destabilization of dipole-mode solitons by partially
incoherent light.
\end{abstract}

\maketitle

Optical vortices are associated with phase dislocations of
diffracting coherent optical beams~\cite{soskin}. When optical
vortices propagate in self-defocusing nonlinear media, the vortex
core with a phase dislocation becomes self-trapped, and the
resulting stationary singular beam is known as an optical vortex
soliton~\cite{swartz,book}. However, in {\em self-focusing
nonlinear media}, optical vortices can exist as ring-like optical
beams carrying a phase singularity~\cite{kruglov} which are known
to be unstable decaying into several fundamental optical
solitons~\cite{firth,book}.

If a vortex-carrying beam is partially coherent, the phase front
topology is not well defined, and statistics are required to
quantify the phase. However, such a partially incoherent vortex
beam can be stabilized in a self-focusing nonlinear medium when
the degree of spatial incoherence exceeds a certain threshold
value, as was recently demonstrated theoretically and
experimentally~\cite{our_prl}.

Waveguides induced by optical vortices in both linear and
nonlinear regimes are of a special interest because this type of
waveguides is robust and can be made
reconfigurable~\cite{wave1,wave2,wave3}. Moreover, the
vortex-induced waveguides can guide large-amplitude beams beyond
the applicability limits of the linear guided-wave theory, and,
together with the guided beam, they can form a {\em vortex-mode
vector soliton} or its dipole-mode
generalization~\cite{vortex,dipole,dipole2}. Recent theoretical
studies, including the rigorous stability analysis~\cite{dipole2},
suggest that the stable propagation of spatial  vortex-like
stationary structures in a self-focusing medium may become
possible in the presence of a large-amplitude beam it guides.

The main purpose of this letter is twofold. First, we demonstrate,
for the first time to our knowledge,  that the initially coherent
vortex beam can guide a partially incoherent light in a
self-focusing nonlinear medium being stabilized by it against the
azimuthal instability and creating {\em a novel type of stable
incoherent soliton}. Second, we demonstrate that in some cases the
incoherence of the guided beam may even enhance, rather than
suppress, the vortex azimuthal instability.

We consider the mutually incoherent interaction of two optical
beams propagating in a self-focusing saturable nonlinear medium
described by the coupled equations,
\begin{equation}
\label{vortex_induced_vaweguides}
   \begin{array}{l} {\displaystyle
      i \frac{\partial u}{\partial z} +\Delta_{\perp} u + F(I_{\rm tot})u = 0,
    } \\*[9pt] {\displaystyle
      i \frac{\partial v}{\partial z} +\Delta_{\perp} v + F(I_{\rm tot}) v = 0,
      }
   \end{array}
\end{equation}
where $u$ and $v$ are the dimensionless amplitudes of two fields,
$F(I) = I/(1+ \sigma I)$ where $\sigma$ characterizes the
nonlinearity saturation effect, and $I_{\rm tot}= |u|^2 + |v|^2$
is the total beam intensity. The spatial coordinate $z$  is the
propagation direction of the beams, and $\Delta_\perp$ stands for
the transversal part of the Laplace operator. The model
(\ref{vortex_induced_vaweguides}) describes interaction of two
mutually incoherent beams in photorefractive nonlinear media when
both anisotropy of nonlinear response and diffusion effects are
neglected. Different types of composite vector solitons in such a
model have been predicted theoretically, and observed
experimentally in photorefractive
crystals~\cite{vortex,dipole,dipole2}.

We consider the case when one of the beams, say $u$, carries a
spatially localized, {\em initially coherent} optical vortex of
the form $u(r, \phi; z) = u(r) \exp(i\phi) \exp(i\beta_1 z)$,
where $\beta_1$ is the vortex propagation constant, the vortex
amplitude function $u(r)$ vanishes for $r \rightarrow \infty$, and
$r$ and $\phi$ are the radius and phase in the cylindrical
coordinates.

When the second field $v$ is also coherent, it can be written in
the form, $v(r,z) = v(r) e^{i\beta_2 z}$, where $v(r)$ is the beam
amplitude and $\beta_2$ is the second propagation constant.
However, when the field $v$ is generated by {\em a partially
incoherent source}, this simple presentation is no longer valid,
and we study the beam propagation numerically employing the
coherent density approach~\cite{IncNum}. This approach is based on
the fact that the partially incoherent field $v$ is presented by a
superposition of mutually incoherent components $v_j$ tilted with
respect to the $z$-axis at different angles, in such a way that
$I_v=\sum_j |v_j|^2$, where $|v_j|^2=G(j\vartheta) I_v$, and
\begin{equation}
G(\theta)= (\pi \theta_0)^{-1/2} \exp(-\theta^2/\theta_0^2)
\label{eq}
\end{equation}
is the angular power spectrum. Thus, coherence of a partially
incoherent light beam is determined by the parameter $\theta_0$,
i.e. less coherence means larger $\theta_0$. Here, $j\vartheta$ is
the angle at which the $j$-th beam in the component $v$ is tilted
with respect to the $z$-axis. For our numerical simulations we
have used a set of 1681 mutually incoherent beams, all initially
tilted at different angles.

\begin{figure}
\centerline{\includegraphics[width=3.1in]{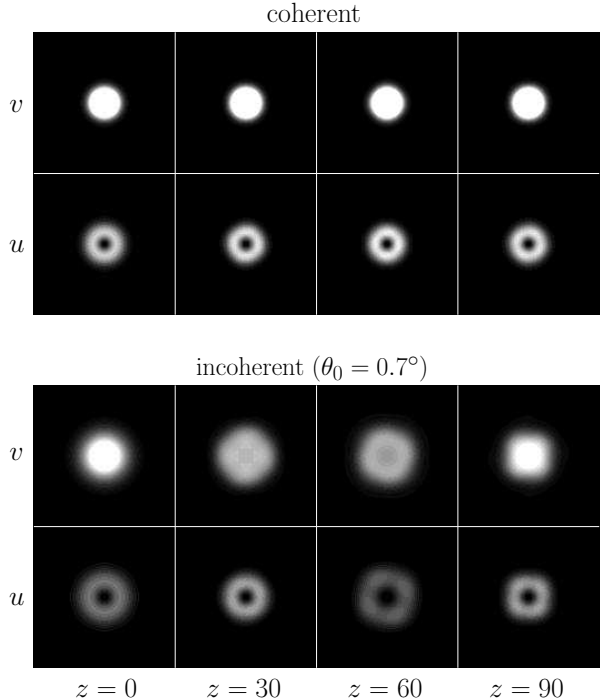}}
\caption{Propagation of the vortex-mode two-component composite
soliton with $\beta_1=1.0$. Upper rows: coherent guided mode with
$\beta_2=1.5$. Lower rows: The same for a partially incoherent
guided mode (at $\theta_0=0.7$); both beams have the same power as
in the coherent case above.} \label{fig1}
\end{figure}

\begin{figure}
\centerline{\includegraphics[width=3.1in]{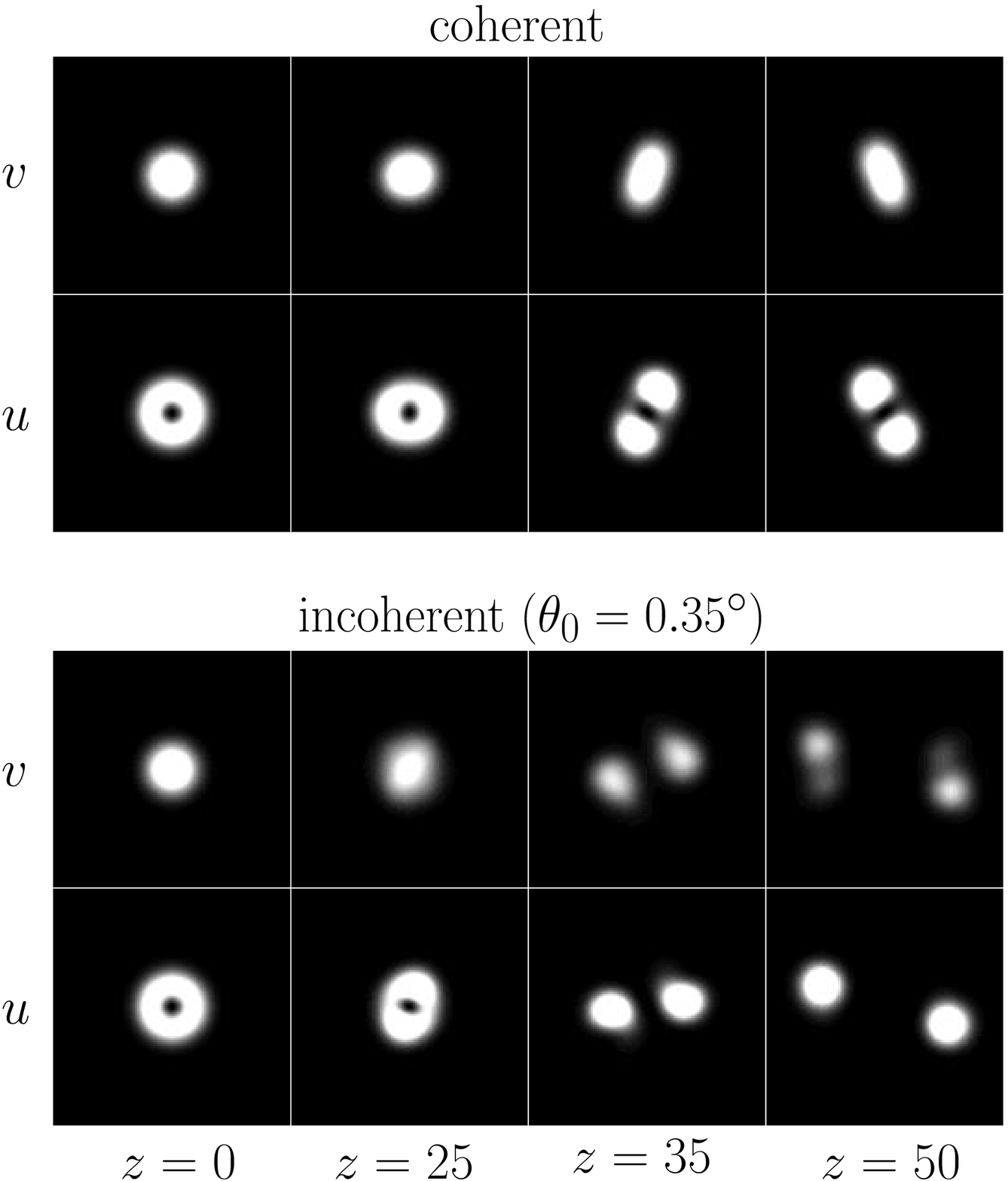}}
\caption{Comparison between the unstable propagation of coherent
and partially incoherent vortex-mode solitons. Upper rows:
coherent vortex at $\beta_1=1.0$ and the coherent guided mode at
$\beta_2=1.45$. The vortex-mode soliton evolves into a rotating
dipole-mode soliton. Lower rows: The same for the partially
incoherent guided mode (at $\theta_0=0.35$); the vortex decays
into two separate beams.} \label{fig2}
\end{figure}

Figure~\ref{fig1} compares the propagation of two-component
composite beams in two cases. In the first case, shown in the
upper two rows of Fig.~\ref{fig1}, the self-trapped vortex $u$ and
the beam $v$ it guides are both {\em coherent}. In a general case,
such a composite beam demonstrates {\em three different scenarios}
of its evolution~(see, e.g., Ref. \cite{dipole2}). When the
amplitude of the guided beam $v$ is small, the vortex $u$ decays
similar to the scalar case~\cite{firth}. For the intermediate
value of the vortex amplitude, the vortex is still unstable but it
evolves into a structure with a rotating dipole component, known
as {\em a dipole-mode vector soliton}~\cite{dipole}. At last, for
relatively large amplitude of the guided beam this {\em composite
partially incoherent vector-mode soliton} becomes stable, see
Fig.~\ref{fig1} (lower two rows).

\begin{figure}
\centerline{\includegraphics[width=3.1in]{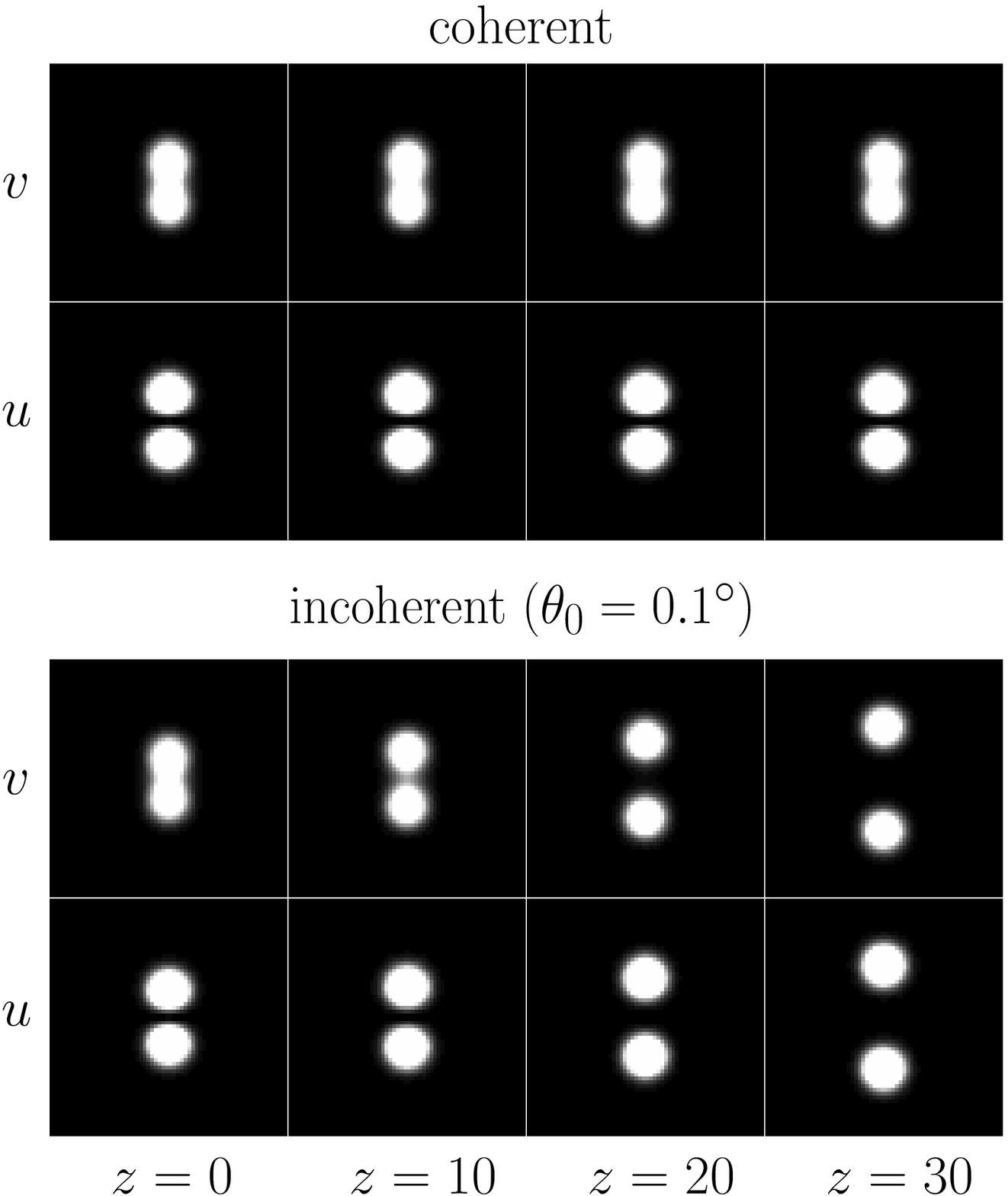}}
\caption{Propagation of the dipole-mode vector solitons with
coherent and incoherent fundamental beams. The initial profile of
the beams corresponds to a solitary solution with propagation
constants $\beta_1=1.0$, for the dipole, and $\beta_2=1.15$, for
the fundamental component. The upper row shows the evolution of
the fundamental, the lower row the dipole. Although the degree of
incoherence is not very high ($\theta_0=0.1^\circ$), it is enough
to destabilize the soliton and leads to its decay. } \label{fig3}
\end{figure}

The mutual interaction between the vortex beam and the mode it
guides has the character of mutual attraction, and it is expected
to provide an effective physical mechanism for stabilizing the
vortex beam in a self-focusing nonlinear medium. Indeed, it is
well-known that the scalar self-trapped vortex beam becomes
unstable in a self-focusing nonlinear medium due to the effect of
the azimuthal modulational instability. In this case, the vortex
splits into fundamental beams that fly off the main vortex
ring~\cite{firth}. On the other hand, bright solitons are known to
be stable in such media. As was demonstrated for two-dimensional
vortex solitons, mutual attraction of the components in a
two-component system may lead to a counter-balance of the vortex
instability by the bright component if the amplitude of the latter
is large enough~\cite{dipole2}.

We study the effect of partial incoherence of the guided mode on
the vortex stabilization. As was mentioned above, for an
intermediate value of the guided-mode amplitude the vortex
structure does not survive and, instead, the vortex transforms
into a dipole-mode soliton~\cite{dipole}. The typical scenario of
such an evolution is presented in Fig.~\ref{fig2} (upper rows).
Due to the initial phase dislocation carried by the vortex, the
resulting dipole rotates during its propagation. However, when the
vortex guides partially incoherent light, we observe that the
resulting dipole soliton {\em becomes more unstable} and, in
particular, the instability of the vortex beam {\em is enhanced}
by the incoherence of the guided mode, as shown in
Fig.~\ref{fig2} (lower rows). The filaments no longer form a rotating
dipole-mode vector soliton, but the filaments fly off the main
vortex ring.

We believe this type of the enhanced instability can be understood
with a simple physical argument. Indeed, the incoherent
fundamental beam can be thought of as many beams that have
different momenta in the transverse plane; these momenta, pointing
away from the center of the beam, add to the momentum of the
vortex beam that decays faster than for the coherent case.

The situation is quite different in the case when the soliton is
stable in the coherent case. Here the incoherence of the
fundamental guided mode seems to have a weak effect on the
propagation of the vortex soliton, and it destabilizes the
composite soliton only very close to the stability threshold and
only when the incoherence is rather strong. Therefore, the
vortex-mode solitons with incoherent fundamental mode show
normally show no sign of instability in  a relatively broad range
of the system parameters (see Fig.~\ref{fig1}).

Thus, partial incoherence destabilizes the rotating dipole-mode
vector soliton that develops from the azimuthal instability of the
vortex. It has also a destabilizing effect on the dipole-mode
vector solitons which are stable in the coherent case. We simulate
the propagation of such solitons, varying the degree of coherence
of the field $v$. An example is presented in Fig.~\ref{fig3}. It
shows the propagation of the dipole-mode soliton with an entirely
coherent fundamental and the propagation of a soliton whose
fundamental is mildly incoherent ($\theta_0=0.1^\circ$). The
fundamental as well as dipole components have equal power in both
cases. It can be seen that the soliton with the incoherent
fundamental component decays whereas the coherent one remains
stable.

%%%%%%%%%%%%%%%%%%%%%%%%%%%%%%%%%

In conclusion, we have introduced a novel type of composite
spatial soliton consisting of a vortex guiding co-propagating
partially incoherent light. The vortex beam, known to be unstable
in a self-focusing nonlinear medium, can be stabilized by a
large-amplitude guided mode above a certain value of its
incoherence, whereas for a low-amplitude bright component the
incoherence may even enhance, rather than suppress, the
instability.

This work was partially supported by the Australian Research
Council and the German Academic Exchange Service (DAAD).
J.R. Salgueiro acknowledges a postdoctoral fellowship of the
Secretar\'{\i}a de Estado de Educaci\'on y Universidades of Spain
partially supported by the European Social Fund.


\begin{thebibliography}{}

\bibitem{soskin} See
M.S. Soskin and M.V. Vasnetsov, in {\em Progress in Optics}, Vol.
42, Ed. E. Wolf (Elsevier, Amstredam, 2001).

\bibitem{swartz} G.A. Swartzlander and C. Law, Phys. Rev. Lett.
{\bf 69}, 2503 (1992).

\bibitem{book} See, e.g., Yu.S. Kivshar and G.P. Agrawal,
{\em Optical Solitons: From Fibers to Photonic Crystals}
(Academic, San Diego, 2003), 560 pp; see Chap. 8.

\bibitem{kruglov} Such beams were first suggested in
V.I. Kruglov and R.A. Vlasov, Phys. Lett. A {\bf
111}, 401 (1985).

\bibitem{firth} W.J. Firth and D.V. Skryabin, Phys. Rev. Lett.
{\bf 79}, 2450 (1997).

\bibitem{our_prl} C.C. Jeng, M. Shih, K. Motzek,
and Yu. Kivshar, Phys. Rev. Lett. {\bf 94}, 043904 (2004).

\bibitem{wave1} A.P. Sheppard and M. Haelterman, Opt. Lett. {\bf 19}, 859 (1993).

\bibitem{wave2} C.T. Law, X. Zhang, and G.A. Swartzlander, Jr., Opt. Lett.
{\bf 25}, 55 (2000).

\bibitem{wave3} A.H. Carlsson, J.N. Malmberg, D. Anderson, M. Lisak
E.A. Ostrovskaya, T.J. Alexander, and Yu.S. Kivshar, Opt. Lett.
{\bf 25}, 660 (2000).

\bibitem{vortex} Z.H. Musslimani, M. Segev, D.N. Christodoulides, and M.
Soljacic´, Phys. Rev. Lett. {\bf 84}, 1164 (2000).

\bibitem{dipole} J.J. Garc\'a-Ripoll, V.M. P\'erez-García, E.A. Ostrovskaya, and Yu.S.
Kivshar, Phys. Rev. Lett. {\bf 85}, 82 (2000).

\bibitem{dipole2} J. Yang and D.E. Pelinovsky, Phys. Rev. E {\bf 67}, 016608 (2003).



\bibitem{IncNum} See, e.g., D.N. Christodoulides, T.H. Coskun, M. Mitchell, and
M. Segev, Phys. Rev. Lett {\bf 78}, 646 (1997).


\end{thebibliography}
\end{document}